\RequirePackage{lineno}
\documentclass[aps,prl,twocolumn,superscriptaddress]{revtex4}
\usepackage{graphicx}
\begin{document}

\title{Measurement of the Cross Section for Prompt Isolated Diphoton
       Production in $p\overline{p}$ Collisions at $\sqrt{s}=1.96$ TeV}

\affiliation{Institute of Physics, Academia Sinica, Taipei, Taiwan 11529, Republic of China} 
\affiliation{Argonne National Laboratory, Argonne, Illinois 60439, USA} 
\affiliation{University of Athens, 157 71 Athens, Greece} 
\affiliation{Institut de Fisica d'Altes Energies, ICREA, Universitat Autonoma de Barcelona, E-08193, Bellaterra (Barcelona), Spain} 
\affiliation{Baylor University, Waco, Texas 76798, USA} 
\affiliation{Istituto Nazionale di Fisica Nucleare Bologna, $^z$University of Bologna, I-40127 Bologna, Italy} 
\affiliation{University of California, Davis, Davis, California 95616, USA} 
\affiliation{University of California, Los Angeles, Los Angeles, California 90024, USA} 
\affiliation{Instituto de Fisica de Cantabria, CSIC-University of Cantabria, 39005 Santander, Spain} 
\affiliation{Carnegie Mellon University, Pittsburgh, Pennsylvania 15213, USA} 
\affiliation{Enrico Fermi Institute, University of Chicago, Chicago, Illinois 60637, USA}
\affiliation{Comenius University, 842 48 Bratislava, Slovakia; Institute of Experimental Physics, 040 01 Kosice, Slovakia} 
\affiliation{Joint Institute for Nuclear Research, RU-141980 Dubna, Russia} 
\affiliation{Duke University, Durham, North Carolina 27708, USA} 
\affiliation{Fermi National Accelerator Laboratory, Batavia, Illinois 60510, USA} 
\affiliation{University of Florida, Gainesville, Florida 32611, USA} 
\affiliation{Laboratori Nazionali di Frascati, Istituto Nazionale di Fisica Nucleare, I-00044 Frascati, Italy} 
\affiliation{University of Geneva, CH-1211 Geneva 4, Switzerland} 
\affiliation{Glasgow University, Glasgow G12 8QQ, United Kingdom} 
\affiliation{Harvard University, Cambridge, Massachusetts 02138, USA} 
\affiliation{Division of High Energy Physics, Department of Physics, University of Helsinki and Helsinki Institute of Physics, FIN-00014, Helsinki, Finland} 
\affiliation{University of Illinois, Urbana, Illinois 61801, USA} 
\affiliation{The Johns Hopkins University, Baltimore, Maryland 21218, USA} 
\affiliation{Institut f\"{u}r Experimentelle Kernphysik, Karlsruhe Institute of Technology, D-76131 Karlsruhe, Germany} 
\affiliation{Center for High Energy Physics: Kyungpook National University, Daegu 702-701, Korea; Seoul National University, Seoul 151-742, Korea; Sungkyunkwan University, Suwon 440-746, Korea; Korea Institute of Science and Technology Information, Daejeon 305-806, Korea; Chonnam National University, Gwangju 500-757, Korea; Chonbuk National University, Jeonju 561-756, Korea} 
\affiliation{Ernest Orlando Lawrence Berkeley National Laboratory, Berkeley, California 94720, USA} 
\affiliation{University of Liverpool, Liverpool L69 7ZE, United Kingdom} 
\affiliation{University College London, London WC1E 6BT, United Kingdom} 
\affiliation{Centro de Investigaciones Energeticas Medioambientales y Tecnologicas, E-28040 Madrid, Spain} 
\affiliation{Massachusetts Institute of Technology, Cambridge, Massachusetts 02139, USA} 
\affiliation{Institute of Particle Physics: McGill University, Montr\'{e}al, Qu\'{e}bec, Canada H3A~2T8; Simon Fraser University, Burnaby, British Columbia, Canada V5A~1S6; University of Toronto, Toronto, Ontario, Canada M5S~1A7; and TRIUMF, Vancouver, British Columbia, Canada V6T~2A3} 
\affiliation{University of Michigan, Ann Arbor, Michigan 48109, USA} 
\affiliation{Michigan State University, East Lansing, Michigan 48824, USA}
\affiliation{Institution for Theoretical and Experimental Physics, ITEP, Moscow 117259, Russia}
\affiliation{University of New Mexico, Albuquerque, New Mexico 87131, USA} 
\affiliation{Northwestern University, Evanston, Illinois 60208, USA} 
\affiliation{The Ohio State University, Columbus, Ohio 43210, USA} 
\affiliation{Okayama University, Okayama 700-8530, Japan} 
\affiliation{Osaka City University, Osaka 588, Japan} 
\affiliation{University of Oxford, Oxford OX1 3RH, United Kingdom} 
\affiliation{Istituto Nazionale di Fisica Nucleare, Sezione di Padova-Trento, $^{aa}$University of Padova, I-35131 Padova, Italy} 
\affiliation{LPNHE, Universite Pierre et Marie Curie/IN2P3-CNRS, UMR7585, Paris, F-75252 France} 
\affiliation{University of Pennsylvania, Philadelphia, Pennsylvania 19104, USA}
\affiliation{Istituto Nazionale di Fisica Nucleare Pisa, $^{bb}$University of Pisa, $^{cc}$University of Siena and $^{dd}$Scuola Normale Superiore, I-56127 Pisa, Italy} 
\affiliation{University of Pittsburgh, Pittsburgh, Pennsylvania 15260, USA} 
\affiliation{Purdue University, West Lafayette, Indiana 47907, USA} 
\affiliation{University of Rochester, Rochester, New York 14627, USA} 
\affiliation{The Rockefeller University, New York, New York 10065, USA} 
\affiliation{Istituto Nazionale di Fisica Nucleare, Sezione di Roma 1, $^{ee}$Sapienza Universit\`{a} di Roma, I-00185 Roma, Italy} 

\affiliation{Rutgers University, Piscataway, New Jersey 08855, USA} 
\affiliation{Texas A\&M University, College Station, Texas 77843, USA} 
\affiliation{Istituto Nazionale di Fisica Nucleare Trieste/Udine, I-34100 Trieste, $^{ff}$University of Trieste/Udine, I-33100 Udine, Italy} 
\affiliation{University of Tsukuba, Tsukuba, Ibaraki 305, Japan} 
\affiliation{Tufts University, Medford, Massachusetts 02155, USA} 
\affiliation{University of Virginia, Charlottesville, VA  22906, USA}
\affiliation{Waseda University, Tokyo 169, Japan} 
\affiliation{Wayne State University, Detroit, Michigan 48201, USA} 
\affiliation{University of Wisconsin, Madison, Wisconsin 53706, USA} 
\affiliation{Yale University, New Haven, Connecticut 06520, USA} 
\author{T.~Aaltonen}
\affiliation{Division of High Energy Physics, Department of Physics, University of Helsinki and Helsinki Institute of Physics, FIN-00014, Helsinki, Finland}
\author{B.~\'{A}lvarez~Gonz\'{a}lez$^v$}
\affiliation{Instituto de Fisica de Cantabria, CSIC-University of Cantabria, 39005 Santander, Spain}
\author{S.~Amerio}
\affiliation{Istituto Nazionale di Fisica Nucleare, Sezione di Padova-Trento, $^{aa}$University of Padova, I-35131 Padova, Italy} 

\author{D.~Amidei}
\affiliation{University of Michigan, Ann Arbor, Michigan 48109, USA}
\author{A.~Anastassov}
\affiliation{Northwestern University, Evanston, Illinois 60208, USA}
\author{A.~Annovi}
\affiliation{Laboratori Nazionali di Frascati, Istituto Nazionale di Fisica Nucleare, I-00044 Frascati, Italy}
\author{J.~Antos}
\affiliation{Comenius University, 842 48 Bratislava, Slovakia; Institute of Experimental Physics, 040 01 Kosice, Slovakia}
\author{G.~Apollinari}
\affiliation{Fermi National Accelerator Laboratory, Batavia, Illinois 60510, USA}
\author{J.A.~Appel}
\affiliation{Fermi National Accelerator Laboratory, Batavia, Illinois 60510, USA}
\author{A.~Apresyan}
\affiliation{Purdue University, West Lafayette, Indiana 47907, USA}
% ...
\collaboration{CDF Collaboration\footnote{With visitors from $^a$University of Massachusetts Amherst, Amherst, Massachusetts 01003,
$^b$Istituto Nazionale di Fisica Nucleare, Sezione di Cagliari, 09042 Monserrato (Cagliari), Italy,
$^c$University of California Irvine, Irvine, CA  92697, 
$^d$University of California Santa Barbara, Santa Barbara, CA 93106
$^e$University of California Santa Cruz, Santa Cruz, CA  95064,
$^f$CERN,CH-1211 Geneva, Switzerland,
$^g$Cornell University, Ithaca, NY  14853, 
$^h$University of Cyprus, Nicosia CY-1678, Cyprus, 
$^i$University College Dublin, Dublin 4, Ireland,
$^j$University of Fukui, Fukui City, Fukui Prefecture, Japan 910-0017,
$^k$Universidad Iberoamericana, Mexico D.F., Mexico,
$^l$Iowa State University, Ames, IA  50011,
$^m$University of Iowa, Iowa City, IA  52242,
$^n$Kinki University, Higashi-Osaka City, Japan 577-8502,
$^o$Kansas State University, Manhattan, KS 66506,
$^p$University of Manchester, Manchester M13 9PL, England,
$^q$Queen Mary, University of London, London, E1 4NS, England,
$^r$Muons, Inc., Batavia, IL 60510,
$^s$Nagasaki Institute of Applied Science, Nagasaki, Japan, 
$^t$National Research Nuclear University, Moscow, Russia,
$^u$University of Notre Dame, Notre Dame, IN 46556,
$^v$Universidad de Oviedo, E-33007 Oviedo, Spain, 
$^w$Texas Tech University, Lubbock, TX  79609, 
$^x$Universidad Tecnica Federico Santa Maria, 110v Valparaiso, Chile,
$^y$Yarmouk University, Irbid 211-63, Jordan,
$^{gg}$On leave from J.~Stefan Institute, Ljubljana, Slovenia. 
}}
\noaffiliation

\date{\today}

\begin{abstract}
This letter reports a measurement of the cross section of prompt isolated
photon pair production in $p\overline{p}$ collisions at a total energy
$\sqrt{s}=1.96$ TeV using data of 5.36 fb$^{-1}$ integrated luminosity
collected with the CDF II detector at the Fermilab Tevatron. The measured
cross section, differential in basic kinematic variables, is compared with
three perturbative QCD predictions, a leading order (LO) parton shower
calculation and two next-to-leading order (NLO) calculations. The NLO
calculations reproduce most aspects of the data. By including photon
radiation from quarks before and after hard scattering, the parton shower
prediction becomes competitive with the NLO predictions.
\end{abstract}

\maketitle

The production of prompt photon pairs with large invariant mass in hadron
collisions is a large irreducible background in searches for a low mass
Higgs boson decaying into a photon pair \cite{higgs}, as well as in searches
for new phenomena, such as new heavy resonances \cite{heavy}, extra spatial
dimensions \cite{dim,grav} or cascade decays of heavy new particles
\cite{casc}. Precise measurements of the diphoton production differential
cross sections for various kinematic variables and their theoretical
understanding are thus very important for these searches. Diphoton
production is also used to check the validity of perturbative quantum
chromodynamics (pQCD) and soft-gluon resummation methods implemented
in theoretical calculations. Diphotons are expected to be dominantly
produced by quark-antiquark annihilation
$q\overline{q}\rightarrow\gamma\gamma$ and also in kinematic regions with
high gluon luminosity, especially at low invariant mass, by gluon-gluon
fusion $gg\rightarrow\gamma\gamma$ through a quark loop diagram. Prompt
photons may also result from quark fragmentations in hard scattering,
although a strict photon isolation requirement significantly reduces the
fragmentation contributions.

Diphoton measurements have been previously conducted at fixed-target
\cite{fixtgt} and collider experiments \cite{ua1,ua2,cdfdip,d0}. The most
recent measurements \cite{cdfdip,d0} were compared with the same pQCD
calculations examined in the present work and large discrepancies were
found between the data and a LO matrix element calculation supplemented
with a parton shower model, suitable for simulation of the backgrounds in
searches of a low mass Higgs boson and of new phenomena. This work shows
that the inclusion of photons radiated from initial and final state quarks
drastically improves the comparison of the parton shower calculation with
the data.

The reported measurement was conducted using data of total integrated
luminosity 5.36 fb$^{-1}$ collected with the Collider Detector at Fermilab
(CDF) \cite{cdf} at the Tevatron $p\overline{p}$ collider. CDF is composed
of a central spectrometer inside a 1.4 T magnetic field, surrounded by
electromagnetic and hadronic calorimeters and muon detection chambers. The
inner spectrometer measures charged particle tracks with a transverse
momentum ($p_{\rm T}$) precision of
$\Delta p_{\rm T}/p_{\rm T}^{2}=0.07\%({\rm GeV/c})^{-1}$. The central
calorimeters cover the region $\vert\eta\vert<1.1$, with an electromagnetic
(hadronic) energy resolution of
$\sigma(E_{\rm T})/E_{\rm T}=13.5\%/\sqrt{E_{\rm T}({\rm GeV})}\oplus 1.5\%$
($\sigma(E_{\rm T})/E_{\rm T}=50\%/\sqrt{E_{\rm T}({\rm GeV})}\oplus 3\%$) and a
tower segmentation of $\Delta\eta\times\Delta\phi\simeq 0.1\times 15^{\circ}$.
Photons are reconstructed in clusters of up to three towers \cite{phosel}.
$\chi^{2}$ criteria are imposed on the profile of the shower to match
expected patterns. Two main cuts are applied: (i) the photon transverse
energy is required to be $E_{\rm T}\ge 17$ GeV for the first photon in the
event and $E_{\rm T}\ge 15$ GeV for the second photon; (ii) the calorimeter
isolation energy in the isolation cone around each photon \cite{caliso} is
required to be less than 2 GeV.

The background from $\gamma$$+$jet and dijet events, where one or two
jets are faking a photon, is subtracted with a method using the track
isolation as the discriminant between signal and background \cite{trkiso}.
It is based on the substantial difference of the track isolation
distribution for signal photons (nearly exponential) and for background
photons (nearly flat). The advantages of this method are that (i) it has
little sensitivity to multiple interactions in the colliding beams, so
that the signal-background separation does not degrade at high instantaneous
luminosity, and (ii) it has high efficiency and good track momentum
resolution, implying minimal degradation of the signal-background separation
due to instrumental effects. The signal fraction is determined by summing
the probabilities of an event to be pure signal, pure background, or mixed
photon pair. These probabilities are obtained by solving a $4\times 4$ matrix
equation using the observation value (0 or 1) for all four combinations
of the leading or sub-leading photon having track isolation below or above
1 GeV/c as an input. The matrix is constructed from the $E_{\rm T}$-dependent
efficiencies of signal and background photons passing the track isolation cut.
A threshold cut of 1 GeV/c is determined by maximizing the separation between
signal and background. The efficiencies are determined from Monte Carlo (MC)
$\gamma$$+$jet and dijet samples, which are produced using the {\sc pythia}
event generator \cite{pythia}. {\sc pythia} events are fully simulated through
the detector and trigger and are reconstructed with the CDF II simulation and
reconstruction software \cite{cdfsim}. With this matrix technique the full
correlations between the two photons in the event are properly taken into
account. Tests are made for underlying event contributions in complementary
cones to the photon reconstruction cone \cite{cones} and also using isolated
tracks in dijet events. The systematic uncertainty in the signal fraction
with this method is of the order of 15-20\%.

The diphoton production cross section differential in a kinematic variable
is obtained from the histogram of the estimated signal in the selected
variable. The average cross section in a bin of the variable is determined
by dividing the bin content by the trigger efficiency, the diphoton selection
efficiency and acceptance, the integrated luminosity and the bin size. The
diphoton trigger efficiency is derived from data \cite{higgs}. It is
consistent with 100\% over all of the kinematic range with a flat
uncertainty of 3\%. The selection efficiency is determined from data and
MC with an iterative method. In the first pass the efficiency is determined
from a fully simulated and reconstructed {\sc pythia} diphoton MC sample by
dividing the number of events passing all selection cuts by the number of
events passing only the kinematic cuts on the photon $E_{\rm T}$, $\eta$,
angular separation and isolation at the event generation level. The
efficiency denominator is corrected for the ``underlying event'' from
collision remnants which make the efficiency obtained from {\sc pythia}
too high by removing events from the denominator through the isolation cut.
This correction is derived by running {\sc pythia} with and without underlying
event and amounts to a constant factor of 0.88 per event. A flat 6\%
uncertainty in the selection efficiency (3\% per photon) accounts for
possible inaccuracies in the {\sc pythia} model for the underlying event.
The signal events of the data are corrected for the preliminary efficiency.
The data are then used to reweight the {\sc pythia} events and obtain a
more accurate representation of the true diphoton distribution. The
efficiency is determined using the reweighted {\sc pythia} sample and
corrected for luminosity dependence, derived from a comparison of the
vertex multiplicity distribution in data and {\sc pythia} MC
$Z^{0}\rightarrow e^{+}e^{-}$ events. The systematic uncertainty in the
efficiency resulting from the luminosity dependent correction grows linearly
from 1.8\% for $E_{\rm T}\le 40$ GeV to 3\% at $E_{\rm T}=80$ GeV and remains
constant above this point. Finally, a 6\% constant uncertainty comes from
the Tevatron integrated luminosity \cite{lumi}.

The $Z^{0}\rightarrow e^{+}e^{-}$ sample is used for calibration by
applying a ``diphoton-like'' event selection, i.e. by imposing a diphoton
selection with the same trigger but allowing for a track associated with
each of the two electromagnetic objects in the event. The electromagnetic
energy scale in data and MC is corrected by tuning the
$Z^{0}\rightarrow e^{+}e^{-}$ mass peak to the world average \cite{zmass}
and a systematic uncertainty from this correction is estimated to grow
linearly from 0 at $E_{T}\le 40$ GeV up to 1.5\% at $E_{\rm T}=80$ GeV
and remain constant above this point. The difference in the photon
identification efficiency between data and MC is estimated from the
$Z^{0}\rightarrow e^{+}e^{-}$ sample \cite{higgs} and added as a systematic
uncertainty to the measurement. All systematic uncertainties in the cross
section measurement are added in quadrature.

The results of this measurement are compared with three theoretical
calculations: (i) the fixed NLO predictions of the {\sc diphox} program
\cite{diphox} including parton fragmentations into photons \cite{frag},
(ii) the predictions of the {\sc resbos} program \cite{resbos} where the
cross section is accurate to NLO, but also has an analytical initial state
soft gluon resummation, and (iii) the predictions of the {\sc pythia} program
\cite{pythia} which features a realistic representation of the physics events
by including parton showering, Initial (ISR) and Final State Radiation (FSR)
and an underlying event model. Diphoton events were selected from an inclusive
$\gamma$$+$X {\sc pythia} sample (X$=$$\gamma$ or jet), thus including the
$q\bar q\rightarrow\gamma\gamma$ and $gg\rightarrow\gamma\gamma$ processes
(56$\%$) as well as the $q\bar q\rightarrow g\gamma\gamma_{_{\rm ISR}}$,
$gq\rightarrow q\gamma\gamma_{_{\rm ISR}}$ and
$gq\rightarrow q\gamma\gamma_{_{\rm FSR}}$ processes (44$\%$). This type of
calculation effectively resums the cross section for gluon and photon
radiation both in the initial and the final state. All calculations are
subject to the experimental kinematic and isolation cuts. {\sc diphox}
accounts for the $gg\rightarrow\gamma\gamma$ process in LO only. The
predictions of {\sc resbos} are restricted to the invariant mass range
from $2m_{b}=9$ GeV/c$^{2}$ to $2m_{t}=350$ GeV/c$^{2}$, where $m_{b}$ and
$m_{t}$ are the masses of the bottom and top quarks, respectively. NLO
theoretical uncertainties are estimated by varying the fragmentation (in
{\sc diphox} only), renormalization and factorization scales up and down
by a factor of two relative to the default scale $\mu=M/2$ of {\sc diphox}
and $\mu=M$ of {\sc resbos}, and for the NLO PDF uncertainties (in both
{\sc diphox} and {\sc resbos}) by using the 20 CTEQ6.1M eigenvectors
\cite{cteq6m}.

\begin{figure*}
\includegraphics[width=7.5cm]{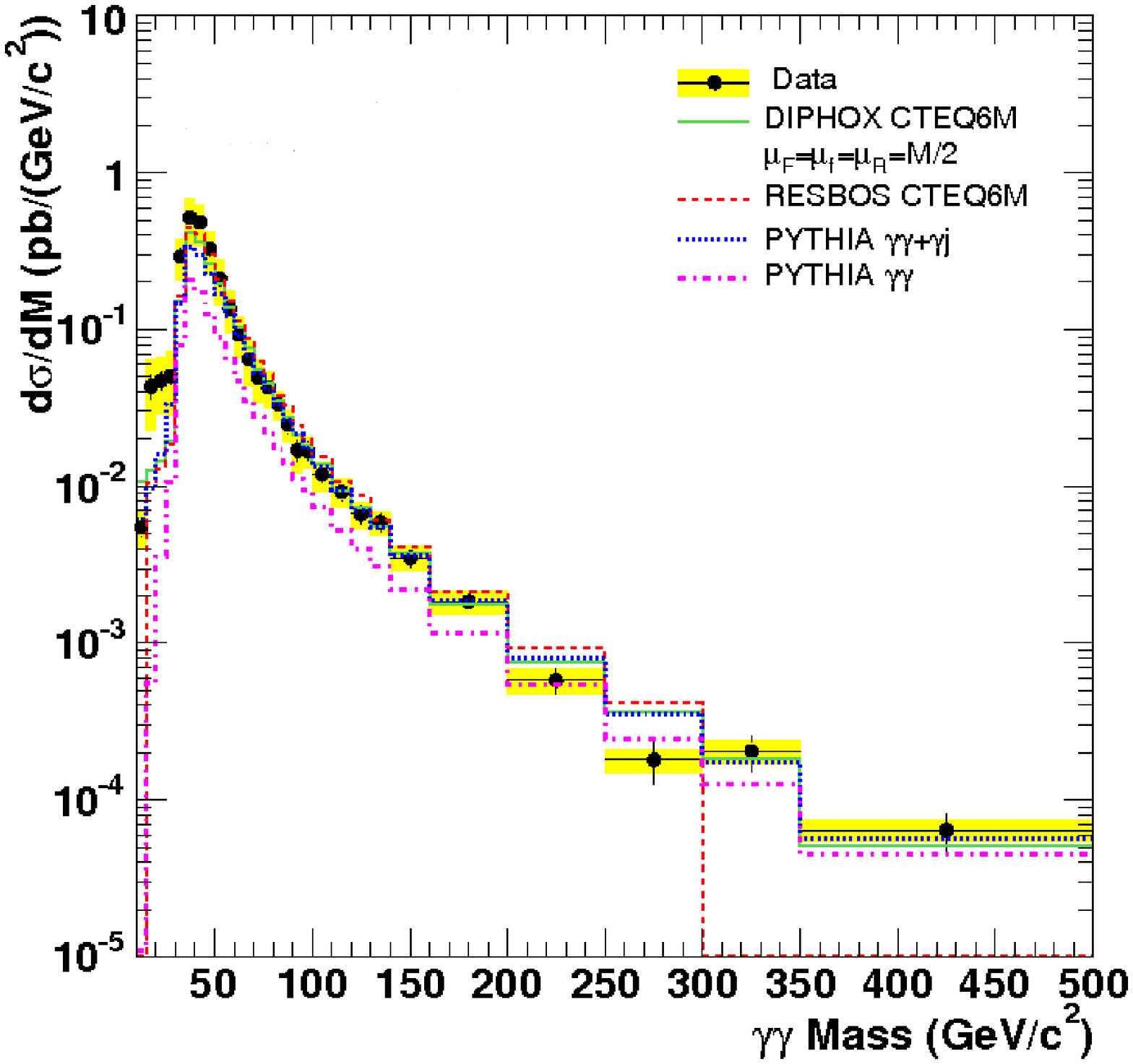}\hspace*{0cm}
\includegraphics[width=7cm]{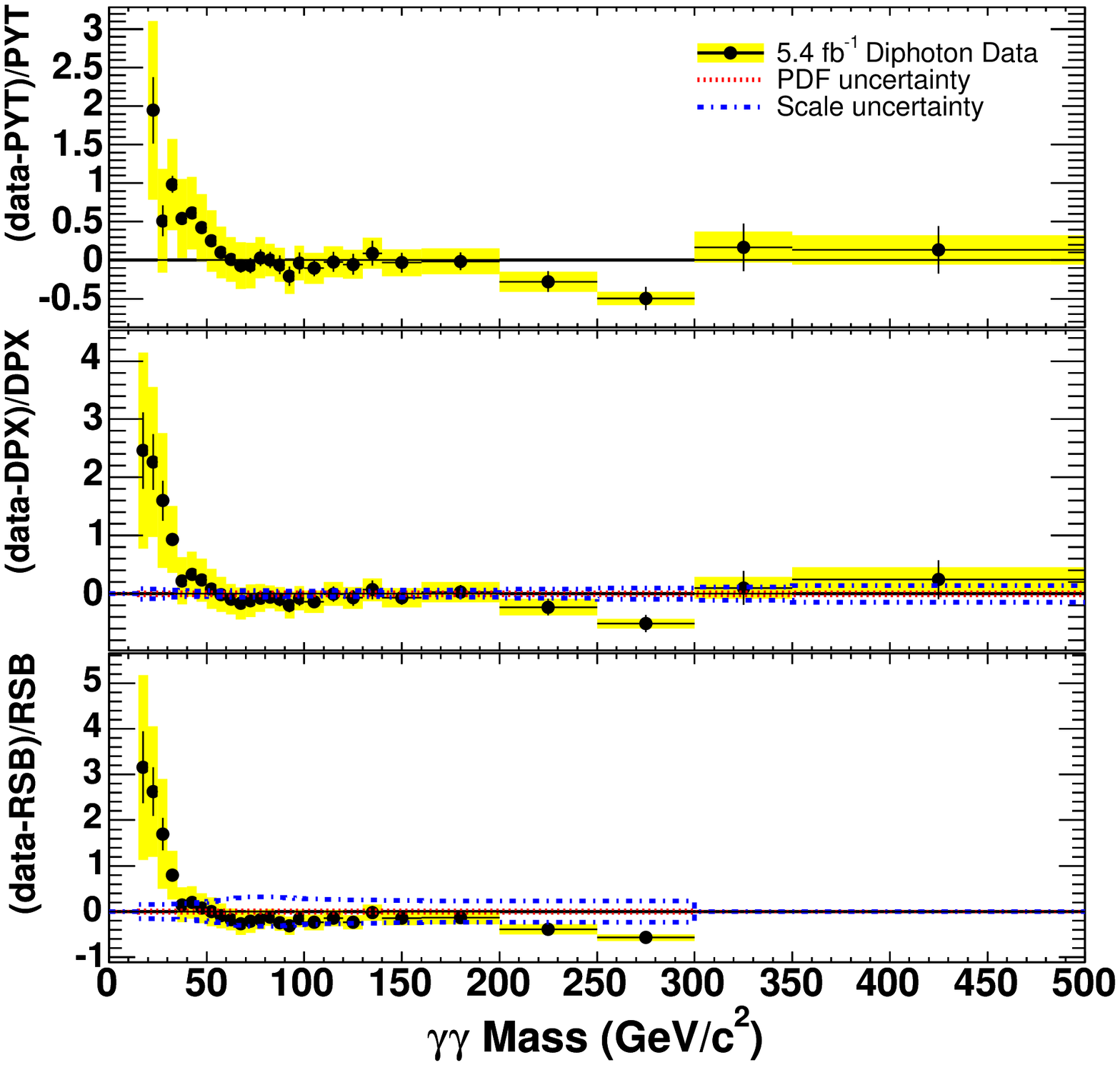}
\vspace*{0cm}
\includegraphics[width=7.5cm]{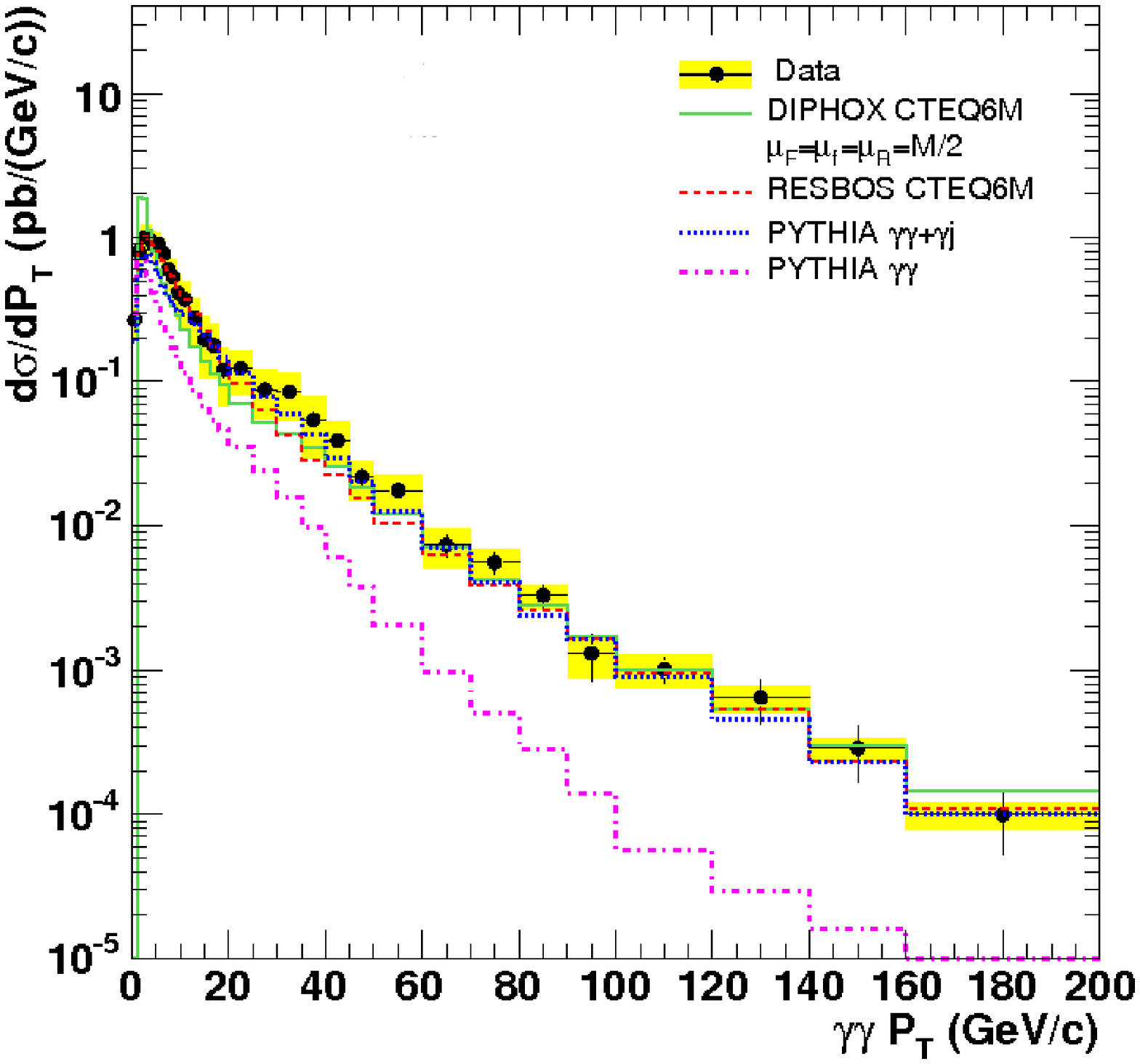}\hspace*{0cm}
\includegraphics[width=7cm]{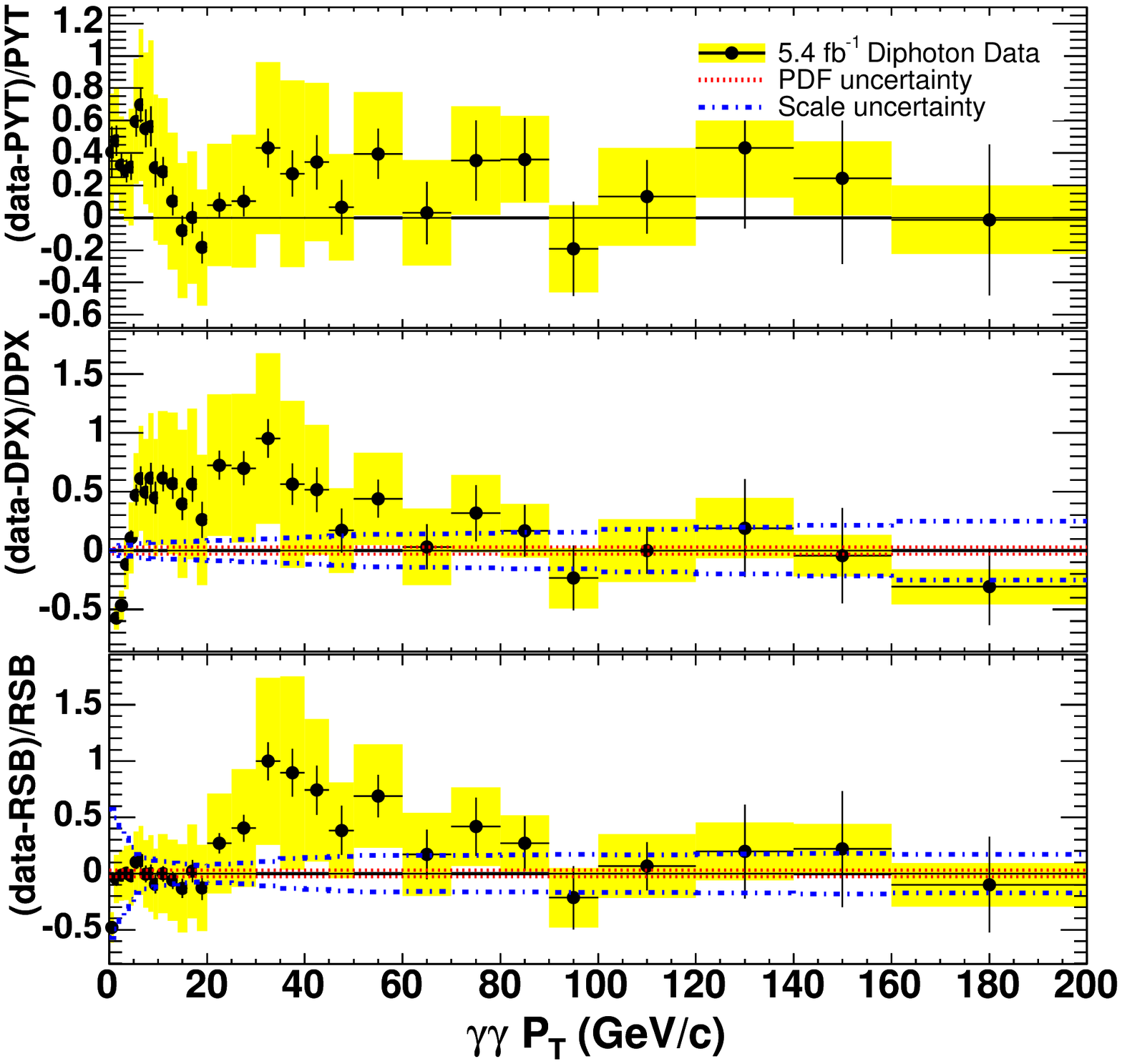}
\vspace*{0cm}
\includegraphics[width=7.5cm]{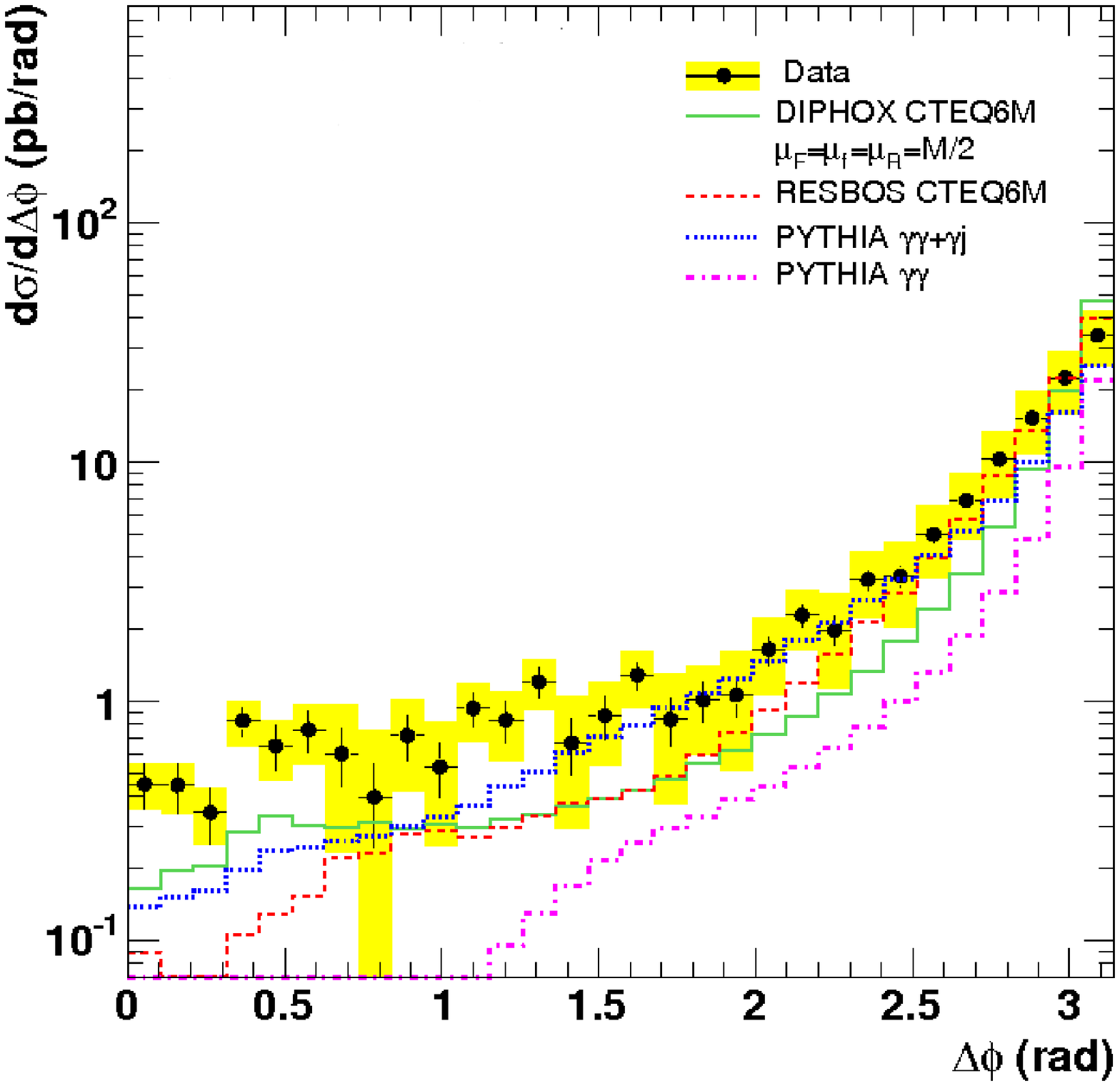}\hspace*{0cm}
\includegraphics[width=7cm]{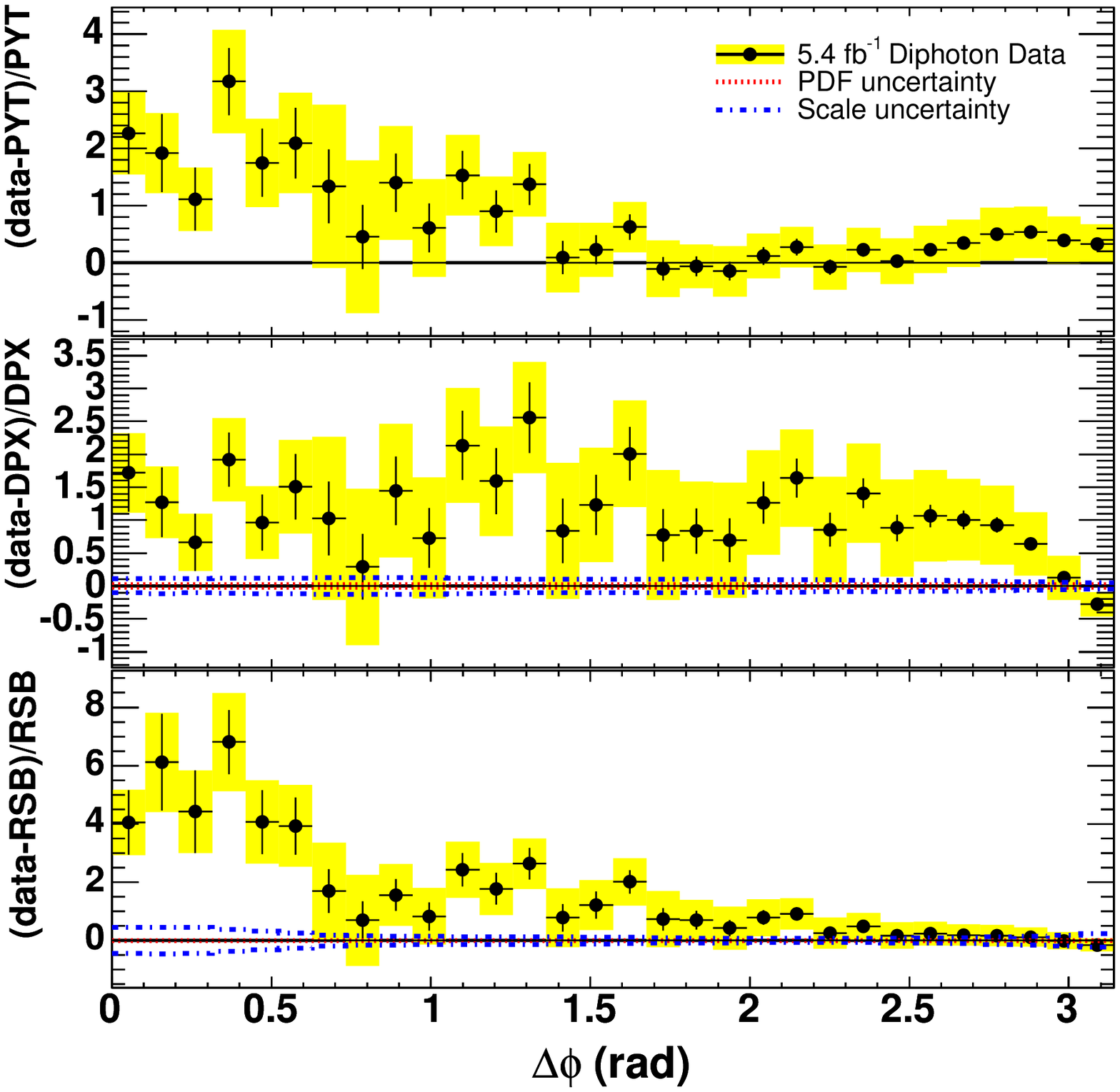}
\vspace*{-0.1cm}
\caption{The measured differential cross sections compared with three
theoretical predictions discussed in the text. The left windows show the
absolute comparisons and the right windows show the fractional deviations
of the data from the theoretical predictions. Fractional deviations for
{\sc pythia} refer to the $\gamma\gamma+\gamma{\rm j}$ calculation. Note
that the vertical axis scales differ between fractional deviation plots.
The comparisons are made as functions of the diphoton mass (top), transverse
momentum (middle) and azimuthal angle difference (bottom). The shaded area
around the data points indicates the total systematic uncertainty of the
measurement.\label{fig:results}}
\end{figure*}

Figure \ref{fig:results} shows the comparison between the measured and
predicted diphoton distributions: the diphoton invariant mass $M$, the
diphoton transverse momentum $P_{\rm T}$ and the difference $\Delta\phi$
between the azimuthal angles of the two photons in the event. While the
{\sc pythia} direct calculation ($\gamma\gamma$) fails to describe both
the scale and shape of the data, including radiation brings the prediction
in fair agreement with the data. In particular, radiation makes the
$P_{\rm T}$ and $\Delta\phi$ distributions harder because of the presence
of at least one hard jet in the final state of events in which one photon
originates from radiation. The mass distributions show a reasonable
agreement with the data for all predictions above the peak at 30 GeV/c$^{2}$,
particularly in the region 80 GeV/c$^{2}$ $<M<$ 150 GeV/c$^{2}$ relevant
to searches for the Higgs boson \cite{higgs}. However, all predictions
underestimate the data around and below the peak. In the $P_{\rm T}$
spectrum all predictions underestimate the data in the region between
20 and 50 GeV/c, a feature also observed in the earlier measurements
\cite{cdfdip,diphox}. For $P_{\rm T}$$<$20 GeV/c, where soft gluon resummation
is most important, only the {\sc resbos} prediction describes the data.
Discrepancies between data and theory are most prominent in the comparison
of the measured and predicted distributions of $\Delta\phi$. In this case
all three predictions fail to describe the data across the whole spectrum.
Approaching $\Delta\phi=\pi$, where soft gluon processes are expected to
manifest, the {\sc resbos} prediction agrees better with the data. In the
range 1.4 rad$<$$\Delta\phi$$<$2.2 rad only the {\sc pythia} prediction
describes the data and remains closest to the data down to 1 rad. In the
low $\Delta\phi$ tail, which corresponds to the region of low $M$
($<$50 GeV/c$^{2}$), all three predictions are lower than the data, although
the {\sc diphox} prediction, by explictly including non-perturbative
fragmentation, lies closer to the data for $\Delta\phi$$<$1 rad.

In summary, the diphoton production cross section, differential in
kinematic variables sensitive to the reaction mechanism, is measured
using data corresponding to an integrated luminosity of 5.36 fb$^{-1}$
collected with the CDF II detector. The high statistics of the measured
sample allows for a higher precision scan over a much more extended phase
space than previous measurements. The overall systematic uncertainty is
limited to about 30\%. The results of the measurement are compared with
three state-of-the-art calculations, applying complementary techniques in
describing the reaction. All three calculations, within their known
limitations, reproduce the main features of the data, but none of them
describes all aspects of the data. The inclusion of photon radiation in the
initial and final states significantly improves the {\sc pythia} parton
shower calculation (see the left-hand panels in Figure \ref{fig:results}),
which is suitable for background simulations in searches for a low mass
Higgs boson and new phenomena.

\begin{acknowledgments}
We thank the Fermilab staff and the technical staffs of the participating
institutions for their vital contributions. We also thank P. Nadolsky, C.-P.
Yuan, Z. Li, J.-P. Guillet, C. Schmidt and S. Mrenna for their valuable help
in the theoretical calculations. This work was supported by the U.S.
Department of Energy and National Science Foundation; the Italian Istituto
Nazionale di Fisica Nucleare; the Ministry of Education, Culture, Sports,
Science and Technology of Japan; the Natural Sciences and Engineering Research
Council of Canada; the National Science Council of the Republic of China; the
Swiss National Science Foundation; the A.P. Sloan Foundation; the
Bundesministerium f\"ur Bildung und Forschung, Germany; the Korean World Class
University Program, the National Research Foundation of Korea; the Science and
Technology Facilities Council and the Royal Society, UK; the Institut National
de Physique Nucleaire et Physique des Particules/CNRS; the Russian Foundation
for Basic Research; the Ministerio de Ciencia e Innovaci\'{o}n, and Programa
Consolider-Ingenio 2010, Spain; the Slovak R\&D Agency; the Academy of
Finland; and the Australian Research Council (ARC). 
\end{acknowledgments}


\begin{thebibliography}{6}
\bibitem{caliso}
The calorimeter isolation is defined as the difference of the transverse
energy $E_{\rm T}=E\sin\theta$ in the isolation cone minus the transverse
energy in the tower cluster of the photon. The isolation cone is defined
to have a radius $R=\sqrt{(\Delta\eta)^{2}+(\Delta\phi)^{2}}=0.4$ around
the axis of the shower profile, where $\eta=-\ln[\tan(\theta/2)]$ is the
pseudo-rapidity, $\theta$ is the polar angle and $\phi$ the azimuth in the
coordinate system of the laboratory with polar axis along the colliding
beams and origin at the center of the detector. The isolation cut implies
that the angular separation of the two photons is $\Delta R\ge 0.4$.
\bibitem{trkiso}
The track isolation is defined as the scalar sum of the transverse momenta
$p_{\rm T}=\vert\vec{p}\vert\sin\theta$ of all tracks originating from the
primary vertex of the event and lying within the photon reconstruction cone.
\bibitem{cones}
The complementary cone is defined by rotating the photon reconstruction
cone by an azimuthal angle of 90$^{\circ}$ about the beam axis, keeping
the polar angle fixed.
\end{thebibliography}
\end{document}